# Velocity and size measurement of droplets from an ultrasonic spray coater using Photon Correlation Spectroscopy and Turbidimetry

PIETER VERDING,[1,2,3] WIM DEFERME,[2,3] AND WERNER STEFFEN[1*]

[1] *Max Planck Institute for Polymer Research, Ackermannweg 10 55128 Mainz, Germany*
[2] *Hasselt University, Instutute for Materials Research (IMO), Wetenschapspark 1, 3590 Diepenbeek, Belgium*
[3] *IMEC vzw, Division IMOMEC, Wetenschapspark 1, 3590 Diepenbeek, Belgium*
*\*Corresponding author: steffen@mpip-mainz.mpg.de*

**Abstract:**

We have developed a combination of light scattering techniques to study and characterize droplets of an ultrasonic spray coater in flight. For this economically relevant spray coater there is so far no reliable technique to systematically adjust the experimental parameters. We have combined Photon Correlation Spectroscopy and Turbidimetry to determine size and speed of the droplets depending on parameters of the printing process as shroud gas pressure, flow rate and atomizing power. Our method will allow to predetermine these parameters to control the properties of the coated films as e.g. thickness from tens of nanometers to micrometers.

## 1. Introduction

In industrial production, especially in the electronic industry, the preparation of thin films with reliable parameters as thickness or roughness, spray coating techniques are indispensable [1-5]. Ultrasonic Spray Coating (USSC) has been introduced for a range of active layers in electronics as organic thin film transistors [3], organic solar cells and photodiodes, electroluminescent devices [6] [7, 8], as well as for electrodes and transistors [3] [6]. Having reliable models [9] and methods to predict vital parameters to achieve reproducible and controlled films is of adamant importance. The droplet size can e.g. be adjusted by regulating the ultrasonic frequency. Because of the small size of the droplets with narrow dispersion in size (less than 20% standard deviation according to [10]), it is possible to coat very thin homogeneous layers down to 10 nm) thickness [11] which is unique compared with other spray coating techniques. This opens up the possibility to deposit functional coatings [12] on large and even three-dimensional surfaces [13]. Another important benefit is that the atomization occurs thanks to high frequency vibrations and no high velocity ejection though a small aperture. This allows the droplets to be deposited with a lower kinetic energy. Since no small aperture is required, the nozzle orifice will be less easily clogged with impurities or solid particles. Lang [14] was one of the first to find a relation between some spray parameters and the droplet size. With the formula of Lang [14] the droplet size can be predicted. For example: spraying water with the impact nozzle of Sonotek at a frequency of 120 kHz, droplets of 18 µm are expected. The relationship found was only correct when the liquid phase viscosity and the volumetric liquid flow rate do not affect the droplet size. R. Rajan, A.B. Pandit and J.Kim [15, 16] have tried to predict the drop size more precisely.

The existing theories do not cover the question of how the droplet size, - velocity, - and concentration evolve during the flight from the ultrasonically generated droplet to the substrate but only focus on the droplet size when it arrives at the substrate. It is, however, important to understand the change of size, velocity and concentration during the flight to predict the influence of the process parameters on the layer formation.

To answer these questions, a measuring technique needs to be developed that can determine the droplet characteristics during flight. For inkjet printing, this led to dimensionless numbers that perfectly describe the ink formulation suitable for printing [17]. However, inkjet printing is jetting only one droplet at the same time and therefore measuring the droplet characteristics is relatively simple. Ultrasonic atomization creates thousands of droplets at the same time. This makes measuring the properties of the droplets as size, distribution in size and velocity during the flight a complicated task. Applying a High-Speed Camera to measure the droplet characteristics during their flight sounds as a valuable solution. Measuring a 3D cone filled with a fog of moving droplets is however not that trivial with a standard High-Speed Camera. To follow the particles frame rates of 20000 are needed due to the speed of the particle; this reduces the numbers of pixels available. Here for a compromise had to be found between number of pixels that represent a droplet and the field of view to track the droplets for the droplet velocity. For our camera and setup this led to 4 pixels for a 20 µm droplet. The error in size is therefore considerable.

We made a survey, which existing techniques could be applied [18] to study the parameters [19] of Ultrasonic Spray Coating. As optical methods we found Fraunhofer diffraction [20], phase doppler anemometry [21], interferometric laser imaging [22], Rainbow refractometry [23], two-phase structured laser illumination planar imaging [24] as possible candidates. The mentioned techniques could be tried on ultrasonic-generated sprays, but this is outside the scope of our work. Our aim is to be able to measure the droplet velocity and size, to make an easy to handle setup. We want to avoid the use of dyes in the inks, as is necessary for some technique above, since this leads to perturbations.

We found that a combination of photon correlation spectroscopy (PCS) (also known as Dynamic Light Scattering (DLS)) to obtain the velocity of the droplets and Turbidimetry to obtain their size is applicable to characterize ultrasonic generated droplets. For the PCS in our case the droplets are the scattering centers instead of particle in solution as one is used to. Due to the high difference in refractive index as well as density between liquid and surrounding gas the droplets lead to a large enough scattering signal. For particle in solution the size of the particle can be obtained from the Brownian motion of these particle. In the case of sprays there is no considerable Brownian motion, thus only the velocity can be determined by PCS and the size has to be obtained from Turbidimetry The droplet size was determined by Turbidimetry, subsequently after PCS but within the same coating step. [18-20]. This requires only the addition of a simple detector to monitor the light intensity passing through the sample.

To corroborate our results, we used a High-Speed Camera for Particle Image Velocimetry/Shadography to see we are in the right order of magnitude of the size and velocity.

## 2. Methods and Theory

### A    Ultrasonic spray nozzle

Schematic setup of an ultrasonic spray nozzle is shown in Fig. 1 with its characteristic parts and parameters. Ultrasonic nozzles [14] produce standing waves in the ink on the atomizing surface. This is the result of mechanical vibrations produced by piezoelectric transducers (Fig. 1.) in the nozzle. The nozzle, powered by a ultrasonic generator delivers an electric signal (parameters: power [W], amplitude $A_e$ and frequency $f_e$ [Hz]) [21] to the cooled transducers. The piezoelectric transducer and the nozzle tip transfer the electric signal to mechanical vibrations ($A_m \propto A_e$, frequency $f_m = f_e$ [22]). Typically used frequencies $f_m$ are located between 25 kHz to 180 kHz. The dimensions of the nozzle are designed to be in resonance in the operating frequency. If the amplitude of the standing waves is high enough, small droplets will break off (atomization effect) [14]. In general, higher frequency nozzles produce smaller droplets [21]. The droplet size is also influenced by adjusting the nozzle height (usually 20-90mm) since the solvent evaporates (partly) during flight. After atomization the droplets do not contain any kinetic energy. Therefore, a so-called shroud gas (e.g.: nitrogen or argon) is used to give the droplets kinetic energy and lead them to the substrate. A higher shroud pressure (psi) will result in droplets with higher velocity. By adjusting the volumetric ink flow rate (ml/min) of the liquid feed, the number of droplets can be strongly influenced. These parameters directly influence the droplet size [23], velocity and the spray angle (α) of the atomized solution. This directly affects the formed layer, especially its thickness.

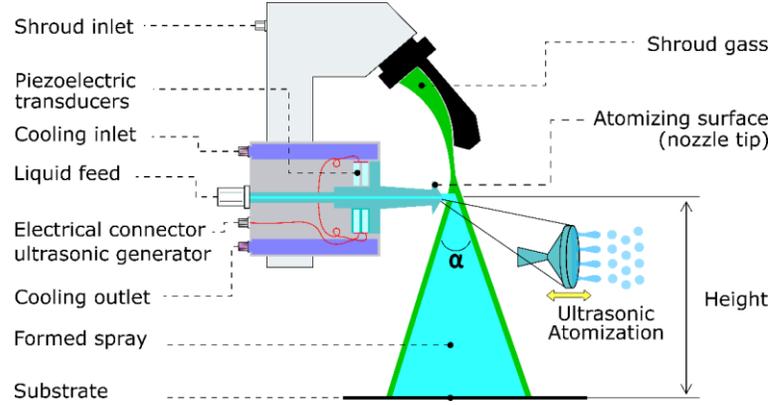

Fig. 1. Ultrasonic nozzle and the shroud gas flow for Ultrasonic Spray Coating

**B    Photon Correlation Spectroscopy**

PCS is a well-seasoned DLS technique [24-26] for the determination of particle sizes via their Brownian motion e.g. in polymer analytics [27]. In the standard experiments, one obtains the intensity-intensity or homodyne autocorrelation function:

$$g_2(q,t) = \langle |E(q,t_0)|^2 |E(q,t_0+t)|^2 \rangle$$
$$\Leftrightarrow \quad g_2(q,t) = \langle I(q,t_0) I(q,t_0+t) \rangle \quad (1)$$

With E the field and I the intensity of the scattered light, q the magnitude of the scattering vector. A representation of the correlation function in terms of a distribution of exponentials, is the so-called Kohlrausch-Williams-Watts function (KWW, Eq. 2) [28, 29] with $\beta$ the stretching parameter describing the distribution[30], A its amplitude. The KWW is in the case of monodisperse particles in solution a pure exponential with $\tau$ the characteristic time at 1/e of this exponential function.

$$g_2(q,t) = A\, exp\left\{-\left(\frac{t}{\tau}\right)^\beta\right\}; \beta = ]0,1] \quad (2)$$

This homodyne correlation function is connected to the field-field or heterodyne autocorrelation function $g_1(q,t)$ via the Siegert relation[26]:

$$g_2(q,t) = C[1 + f|g_1(q,t)|^2] \quad (3)$$

Following [26], $g_1(q,t)$ can experimentally be obtained by adding a local oscillator, i.e. direct laser light into the detector. The dominating term in the expansion of the mathematical representation of such a heterodyne experiment is $g_1(q,t)$ [26]. An effect to note is, that in the case of multiple relaxation processes in the case of $g_2(q,t)$ these are products of (stretched) exponentials while in the case of $g_1(q,t)$ these are a sum. The main advantage of heterodyne light scattering is that for low scattering signals the signal to noise of the correlation functions is enhanced.

Compared to standard PCS experiments there are two questions to address in relation to our experiment:

(a) Is the comparatively low number of scattering droplets problematic? This is not the case: PCS is an interferometric technique in the sense, that one observes the produced speckle field in the far field and its dynamic change. This speckle field is independent of the number of scattering moieties. Modern PCS instruments utilize as in our case single mode detection – we observe only the change of a single speckle since every speckle represents a different electrodynamic mode [31]. This maximizes the amplitude called contrast in PCS and can be used to minimize systematic noise [32].

(b) Is the size of the particles/droplets compared to the wavenumber k of the incident light causing problems in determining size or velocity? For our experimental setup and our aim this is not a problem. One has to take into account not only scattering but refraction and scattering within the droplets. It was found that this influences the results at small angles close to the forward direction [33-36]. We perform PCS at 90° scattering angle.

### *Droplet velocity*

If the particle respective the droplets in a PCS experiment have a directed flow, than due to the Doppler shift one observes in the frequency domain in the spectral density $S(q,\omega)$, a frequency shifted peak. This is in the time domain of PCS a damped oscillation described by a cosine function [37, 38] additionally to the Brownian motion described by a KWW function (Eq. 2):

$$g_1(q,\tau) = A_1 \exp\left\{-\left(\frac{t}{\tau_1}\right)^\beta\right\} + A_2 \exp\left\{-\left(\frac{t}{\tau_2}\right)\right\} \cos(\omega_D t) \quad (4a)$$

$$\overset{FT}{\Leftrightarrow} \quad S(q,\omega) = \left(2\frac{B_1}{\pi}\right)\left(\frac{\Gamma_1}{4(\omega)^2+\Gamma_1^2}\right) + \left(2\frac{B_2}{\pi}\right)\left(\frac{\Gamma_2}{4(\omega-\omega_{c2})^2+\Gamma_2^2}\right) \quad (4b)$$

$A_{1,2}$, $B_{1,2}$ amplitudes, $\tau_{1,2}$ characteristic times, and $\omega_D$ the frequency of the Doppler shift, $\omega_{c2}$ position of the Doppler peak, $\Gamma_{1,2}$ full width half maximum FWHM. $g_1(q,\tau)$ is related to $S(q,\omega)$ via the Fourier-transformation. A central, Lorentzian peak in $S(q,\omega)$ related e.g. to the ubiquitous Brownian motion is transformed into the exponential decay of Eq. 4a, a shifted Lorentzian peak is due to the Doppler shift of the moving particles. However, in the case of ultrasonic spray coating, the particles are in principle droplets that move in and with the shroud gas.

In general, the induced Doppler shift $\omega_D$ is directly related to the droplet velocity:

$$\omega_D(q) = \pm \boldsymbol{q} \cdot \boldsymbol{V} = \pm 2k_i V \cos(\phi) \sin\left(\frac{\theta}{2}\right) \quad (5)$$

The light is scattered into an angle $\theta$, in our case 90°. The angle between the velocity vector **V** and the scattering vector **q** is defined as $\phi$. It should be noted that when **V**⊥**q** the frequency or Doppler shift cannot be observed. To be able to measure we tilted the beam of the shroud gas by 5° against the vertical, now **q** and **V** are not perpendicular to each other anymore. For determining the propagation vector $k_i$ the Bragg condition can be used where q =|**q**| is the magnitude of **q**:

$$q = 2k_i \sin\left(\frac{\theta}{2}\right) = \frac{4\pi n}{\lambda_i} \sin\left(\frac{\theta}{2}\right) \rightarrow k_i = \frac{2\pi n}{\lambda_i} \quad (6)$$

Where n is the refractive index of the measured medium. The incident light wavelength is represented as $\lambda_i$. If Eq. (5) and Eq. (6) are combined the droplet velocity can be extracted.

$$V = \frac{\omega(q)}{2k_i \cos(\phi)\sin\left(\frac{\theta}{2}\right)} = \frac{\omega(q)\lambda_i}{4\pi n \cos(\phi)\sin\left(\frac{\theta}{2}\right)} \quad (7)$$

In the final Eq. (8) $\omega(q)$ is replaced with the frequency (f (q)). The velocity of the droplet can be calculated by having one variable parameter f. The frequency is defined by fitting the correlation function [27] obtained through PCS as will be shown later.

$$V = \frac{f(q)\lambda_i}{2n\cos(\phi)\sin\left(\frac{\theta}{2}\right)} \qquad (8)$$

**C    Turbidimetry**

The droplet size can be determined by performing one additional Turbidimetry measurement [39]. In this experiment, the amount of transmitted light is measured and placed in relation to the droplet diameter followed by the Mie theory. For turbidimetry the advantage of determining size and concentration at the same time using several wavelengths is not straightforward possible, only one of the parameters can be determined independently. The Mie scattering at diameters above 5µm show too little difference (Fig. 2) using different wavelength.

*Droplet size*

The size of droplets in suspension can be determined by measuring the turbidity $\tau_{\lambda 0}$. Turbidimetry measures the damping of a light beam traveling through the spray caused by the absorption and scattering of light by the droplets/particles.

$$\tau_{\lambda 0} = \frac{1}{L}\ln\left(\frac{I_0}{I}\right) \qquad (9)$$

$I_0$ and I are representing the intensities of the incident and weakened light beams. The optical path length through the measured medium is L. The amount of absorption and scattering is related to the concentration and size of the droplets. Therefore, turbidity can be related to the droplet size. For droplets with monodisperse spherical diameter, D, a relation is defined [39]:

$$\tau_{\lambda 0} = \phi_N \frac{\pi D^2}{4} Q_{ext} \qquad (10)$$

Absorption in a medium is related to the extinction coefficient $Q_{ext}$. This depends on the incident light wavelength $\lambda_0$ and droplet diameter D. The concentration of the droplets is defined as $\phi_N$. After combining Eq. (9) and Eq. (10) D can be determined:

$$D = \sqrt{\frac{-\ln\left(\frac{I}{I_0}\right)4}{L\phi_N Q_{ext}\pi}} \qquad (11)$$

The extinction coefficient $Q_{ext}(D)$ has an asymptote of two (Fig. 2). In our experiment $Q_{ext} = 2$ is appropriate to calculate the droplet diameter because the expected average droplet size is 18 µm. In Turbidimetry the particle size and concentration can be determined independently by using a set of different wavelengths. In our experiment there is only a negligible difference in extinction coefficient between two wavelengths for droplets between 10-100 µm. Based on this, only the droplet diameter can be determined with the Turbidimetry measurement. The amount of multiple scattering is in our case limited and negligible, thanks to the light transmission higher than 95% [40] as shown from our experimental data in the inset in figure 2. As was pointed out in [40, 41] only for transmission below 40% or vice versa obscuration/obfuscation 50% a considerable influence of multiple scattering is expected. This high transmission is a direct consequence of the low density of droplets in this technique. This high transmission allows us to perform the high-speed camera observation across the spray cone and the droplets in the focus of the camera are well resolved.

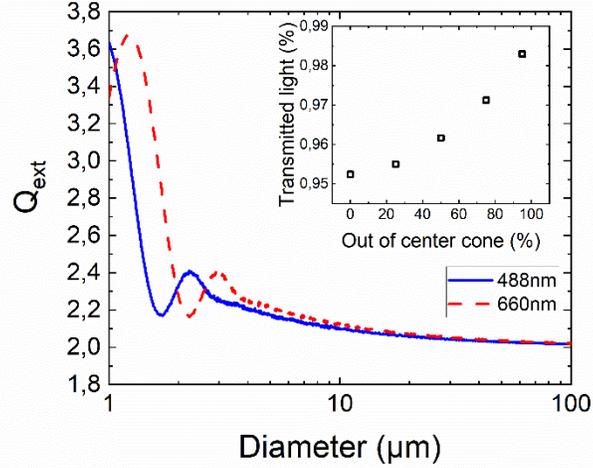

Fig. 2: The diameter of the droplets influences the extinction coefficient ($Q_{ext}$). This is modelled with two different wavelengths of the incident light: 488 nm and 660 nm for droplet size distribution with standard deviation of 20 % for each particular median diameter. (data generated with: http://philiplaven.com/mieplot.htm) The inset shows the measured transmitted light intensity vs. the horizontal place in the spray cone. In this experiment the diameter of the cone at the place the transmission was measured was 34 mm.

To solve Eq. (11), there are two unknown parameters, namely the diameter and the concentration of the droplets. The other parameters are known. To obtain the droplet diameter, Eq. (12) is applied.

$$\phi_N = \frac{\dot{V}_{liquid}}{V_{sphere} A V} \qquad (12)$$

Here, $\dot{V}_{liquid}$ is the volumetric flow rate of the syringe pump. The volume of the droplet $V_{sphere}$ can be replaced by an expression including the diameter D assuming the droplets are perfectly spherical. The velocity $V$ of the droplets will be defined by PCS. Eq. (8) is only correct if one assumes that the velocity over the entire surface A is constant and perpendicular in the cone. Therefore, edge effects of the velocity are not considered in this work. The surface area A can be determined by using a simple camera to measure the angle of the cone. From Eq. (11) and (12), the droplet diameter can be obtained.

### 3. Materials and Setup

For the combination of heterodyne PCS and Turbidimetry, an experiment was developed (Fig. 3) Droplets were created and sprayed from an ultrasonic spray nozzle (Sono-Tek, Impact). This nozzle was mounted on an XYZ-stage under an angle of 85 degrees to the scattering plane.

**Turbidimetry:** A laser diode (Qioptiq/Excelitas, iFLEX2000, λ = 660nm, 40mW) was used as light source, after the laser, a beam expander (X3) was placed. The beam expander is needed to reduce the energy density of the light on the detector to avoid saturation by maximizing the signal. The intensity of the expanded beam can be fine-tuned with an attenuator (neutral density filter). This light transmits through the spray of droplets and reaches the custom build detector (including photodiode 220D, OEC GmbH). In front of the detector, a pinhole is placed whose opening is adjusted to the size of the laser beam diameter. Therefore, only transmitted light enters the detector reducing the amount of stray light. Two flip mirrors made it possible to switch easily from Turbidimetry to the PCS measurement (Fig. 3).

**Photon Correlation Spectroscopy:** Measurements were performed with a heterodyne setup. This means that elastic light and inelastic light are combined for self-beating. The heterodyne detection mode was chosen for its stability and better signal/noise ratio. Light from the blue laser diode ($\lambda$ = 488nm) is divided into two beams by using a parallel glass plate with a reflectance of 5%. The intensities of both beams can be regulated separately with attenuators (Fig. 3). The scattered light coming from the spray falls into a single mode fiber with a Y-beam splitter into two single-photon counting modules (Avalanche diode, Perkin Elmer, SPCM-AQR). This signal is correlated by the Multiple Tau Digital Correlator (ALV GmbH, ALV-7004) in pseudo cross-correlation.

**High Speed Camera:** A High-Speed Camera (Photron, Mini AX100 200K-M-32GB) is mounted together with a Bi-Telecentric objective (Thorlabs, X2, MVTC23200) as a reference measurement system. Illumination of the view field by a white light source (SCHOTT, KL 2500 LCD) combined with a telecentric backlight illuminator (Techspec, 52 mm). A frame rate of 20000 frames/second was needed to obtain the velocities with high enough precision. This leads to a reduction in the field of view for the camera to attain this frame rate. Therefore, in our experiment the resolution for a 20 µm particle is about 4 pixel.

The experimental setup (Fig. 3). was situated on a breadboard with active vibration insulation (Scientific Instruments GmbH, TableStableTS300) mounted in a light-tight black box (eliminating disturbing correlations caused by extraneous scattered light sources as room lighting, sunlight).

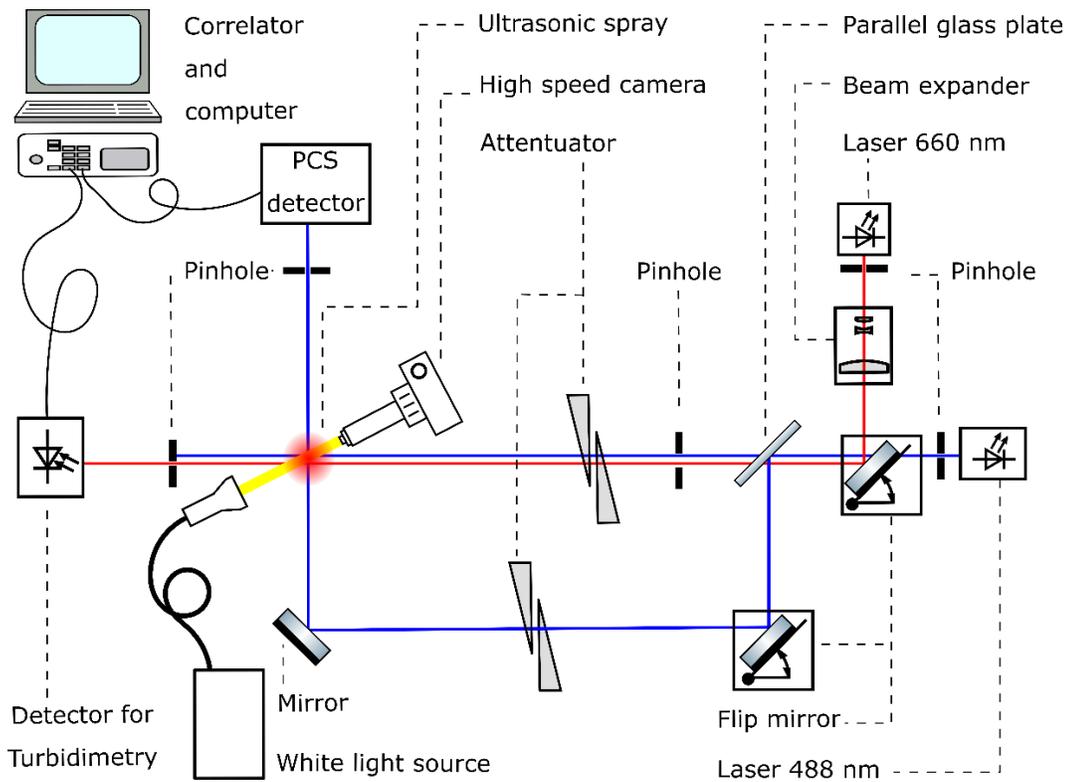

Fig. 3. The experimental setup for Turbidimetry combined with PCS: Three measurement techniques are combined: heterodyne PCS to measure the droplet velocity, Turbidimetry measuring the droplet size and a High Speed Camera to corroborate our results. The ultrasonic spray coater is symbolized by the scattering volume as a red dot and not explicitly drawn for clarity.

## 4. Data Treatment

After obtaining the raw data from the experiments, it is necessary to process this data in order to obtain the required results. In this chapter, it is explained how to transform the obtained data to the needed parameters for the Eq. (8). Subsequently, the droplet velocity-diameter can be obtained.

### Photon Correlation Spectroscopy

A digital multitau correlator (ALV GmbH, ALV-7004) is used for signal processing to obtain $g_1(q,t)$ (Fig. 4) or $g_2(q,t)$. The correlation function itself has an exponential decay with a shifted additional peak. This peak represents the Doppler shift. To obtain the velocity of the droplets, the angular frequency of this peak must be determined. Fig. 4 shows that it is not possible to fit the correlation function properly and extract the angular frequency from the attenuated cosine which is strongly damped.

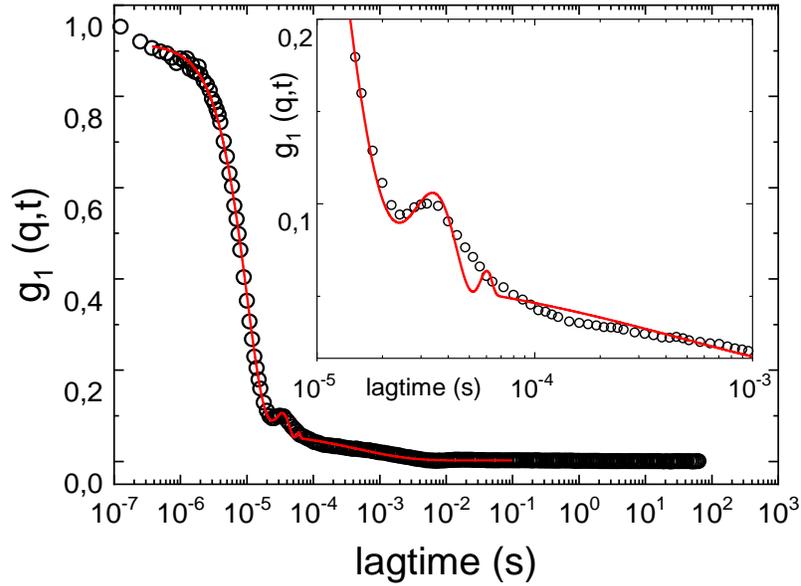

Fig. 4. Heterodyne correlation function $g_1(q,t)$ of a PCS measurement – the small peak at $3 \cdot 10^{-5} s$ contains information about the measured droplet velocity. $g_1(q,t)$ is fitted with Eq.(4a) [26] (function displayed in Table 1). The experimentally observed cosine is strongly damped. The inset shows an enlargement of the main feature. The fit function is displayed as a full (red) line.

**Table 1. The fit function used (Fig. 4). This is a combination of KWW Eq. (2) and Eq. (4a).**

| Equation | |
|---|---|
| | $y = A_1 exp\left[-\left(\frac{t}{\tau_1}\right)^\beta\right] + A_2 exp\left[-\left(\frac{t}{\tau_2}\right)\right][cos(\omega t)]^2 + BG$ |

Therefore, we made a Fast Fourier transform (FFT, Fig. 5) of the data. Now the Doppler shift can be read out from the real part of the spectrum. Modern correlators like the one used, have a quasi-logarithmic time base. A FFT needs a linear time base with the sampling rate to be double as the highest frequency in the data (Niquist theorem). Therefore, before applying the Fourier transformation, the data is interpolated linearly with a time base equivalent to the fastest lag time.

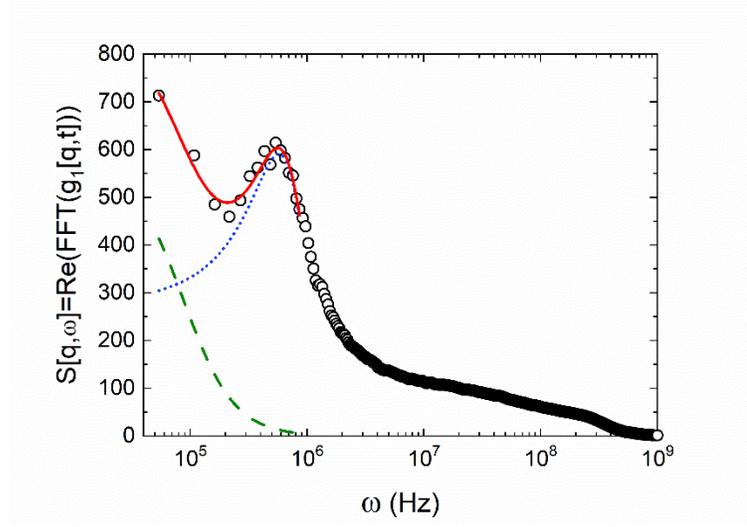

Fig. 5. S(q,ω), the real part of the Fourier transformation of $g_1(q,t)$. The frequency spectrum is fitted with two Lorentzians according to Eq. 4b. The cumulative peak of the fit is displayed on top of the function as a full (red) line.

**Table 2. Parameters of fit with two Lorentzians**

| Equation | $S(q,\omega) = y_0 + \left(2\dfrac{B_1}{\pi}\right)\left(\dfrac{\Gamma_1}{4(\omega)^2 + \Gamma_1^2}\right) + \left(2\dfrac{B_2}{\pi}\right)\left(\dfrac{\Gamma_2}{4(\omega - \omega_{c2})^2 + \Gamma_2^2}\right)$ |
|---|---|

|  |  | Value |
|---|---|---|
| **FWHM1** | $\Gamma_1$ | 1.75E5 |
| **Amplitude1** | $B_1$ | 1.57E8 |
| **Peak2** | $\omega_{c2}$ | 5.74E5 |
| **FWHM2** | $\Gamma_2$ | 1.08E6 |
| **Amplitude2** | $B_2$ | 9.97E8 |
| **baseline** | $y_0$ | 0 |

The shifted peak against the central Lorentzian represents the Doppler shift (Fig. 5). The fit was performed with the Levenberg-Marquart algorithm used in Origin (Originlab.com) non-linear fitting routine. To reduce the complexity, only two Lorentzians are fitted in the interested area. These are the central Lorentzian (dashed green line) and the Lorentzian of the first large peak (dotted blue line). The Doppler shift ($\omega_{c2}$) is obtained after fitting. This frequency (574000 Hz) can be entered in Eq. (8) to calculate the velocity of the droplet. The resulting velocity is in this case 2,273 $m/s$ (with parameters: $f = 574000\ Hz, \lambda = 488 \cdot 10^{-9} m, \phi = 85°$ and $\theta = 90°$).
Different spray parameters influence the shape and placement of the spray.

**Turbidimetry.**

The intensity measurement is performed for a decade of seconds. The average value is calculated, this number is entered directly in Eq. (11). Together with the velocity of the droplets from the PCS measurement, Eq. (12) can be applied to determine the size of the droplets.

## 5. Results and discussion

To test our measurement system, we performed USSC experiments with water as ink. Here, the shroud pressure, volumetric flow rate of the ink and atomizing power have been increased to study the droplet size and velocity. The results are compared with the data from the High Speed Camera as well as with predictions from theory for the droplet diameter [16]. Theoretical calculations or initial experimental measurements for the droplet speed for USSC are not available in literature, indicating the innovative character of this research.

### 1. *Increasing shroud pressure*

#### a) **Droplet velocity**

The frequency shift was studied increasing the shroud pressure (parameters Table 3). If the shroud pressure increases, so should the droplet velocity. By increasing the shroud pressure the second peak shifts to the right (Fig. 6). Meaning that the second peak is located at a higher frequency resulting in a bigger Doppler shift giving a higher measured droplet velocity (Eq. 8). This proves that PCS can detect velocity changes in ultrasonic sprays.

Table 3 . Spraying parameters of the series of experiments in Fig. 6 with increasing shroud pressure.

| Shroud pressure (psi) | Flow rate (ml/min) | Atomizing power (W) | Spray height (mm) | Ink |
|---|---|---|---|---|
| 0.2-0.4-0.6-0.8-1.0-1.2-1.5 | 1.5 | 2.5 | 40 | Water |

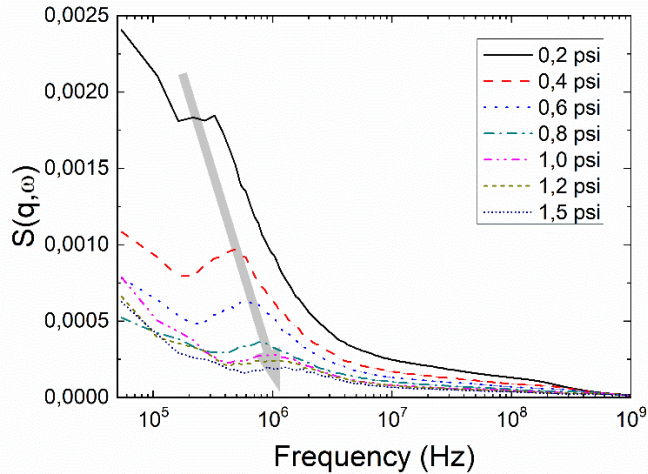

Fig. 6: Fast Fourier transformation of the experimentally obtained $g_1(q,t)$. The peak of the second Lorentzian shifts to the right by increasing the shroud pressure resulting in a higher droplet velocity.

Three independent series of measurements (Table 3) where performed. The obtained droplet velocity from the PCS and HSC correlate (Fig. 7 a and b). The results of the PCS have the same magnitude and trend as measured with the HSC. The droplet velocity increases with increasing shroud pressure. The spread and accuracy of the results are improved with the PCS compared to the HSC.

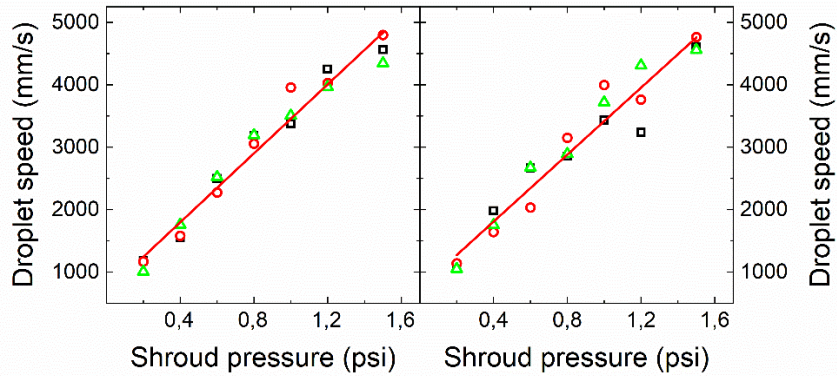

Fig. 7 . Droplet speed vs. increasing shroud pressure (pressure of the nitrogen shroud gas) from three independent runs; line is the average as linear fit. The results of the PCS (a) have the same slope within the error as the HSC (b). Higher accuracy of PCS results compared to the HSC (Slope PCS:2774 and HSC 2683).

**Droplet diameter**

Turbidimetry measurements were made as a series of three independent experiments (Table 3). Using the from PCS obtained droplet velocity (Fig. 7a), the average diameter of the droplets is obtained (Fig. 8). Spraying water with the impact nozzle (120 kHz), the expected droplet size according to Lang [14] agreed with our Turbidimetry measurement. The average diameter is $18 \pm 3$ µm (Fig. 8) and is not influenced by the shroud pressure.

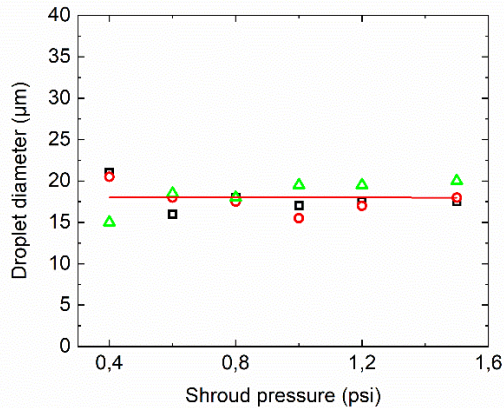

Fig. 8 . Droplet diameter vs. increasing shroud pressure (pressure of the nitrogen shroud gas) from three independent runs; line is the average as linear fit. – no influence in the shroud pressure range studied (slope: -0,03).

*Increasing the volumetric ink flow rate*

**Droplet velocity**

The effect of the volumetric ink flow rate on the droplet velocity was investigated with PCS by performing three independent series of measurements (Table 4). Decreasing droplet velocity is observed with increased volumetric ink flow rate (Fig. 9a). The magnitude and trend did agree with the HSC (Fig. 9b).

Table 4. Spraying parameters of the series of experiments in Fig. 9 with increasing ink flow rate.

| Shroud pressure (psi) | Flow rate (ml/min) | Atomizing power (W) | Spray height (mm) | Ink |
|---|---|---|---|---|
| 0.8 | 1.0-2.0-3.0-4.0 | 2.5 | 40 | Water |

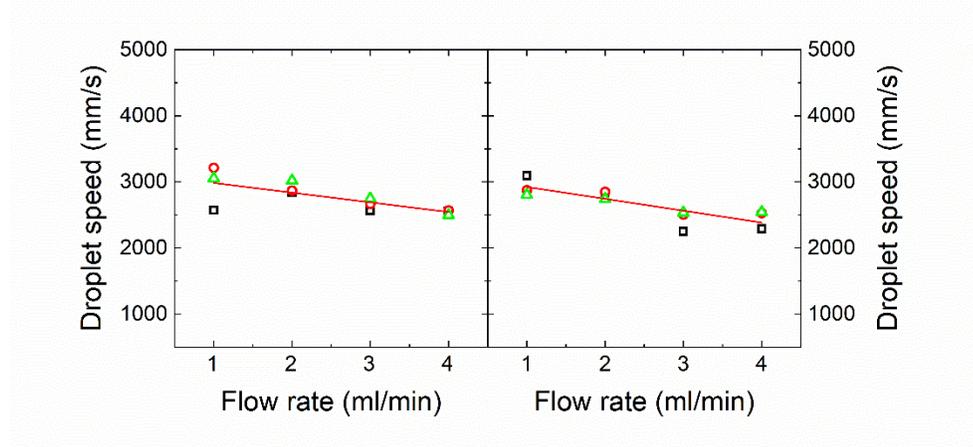

Fig. 9 . Droplet speed vs. increasing volumetric ink flow rate. (a) PCS, (b) HSC show the same trend and quantitative results; full (red) line is the average as linear fit (Slope PCS: -147 and HSC -180).

**Droplet diameter**

Turbidimetry measurements were made from three series of independent measurements (Table 4). Together with the obtained droplet velocity (Fig. 9a) and the Turbidimetry measurement the droplet diameter is obtained (Fig. 10). This diameter is dependent on the volumetric ink flow rate and did agree with the theoretical predictions according to Rajan and Pandit [16].

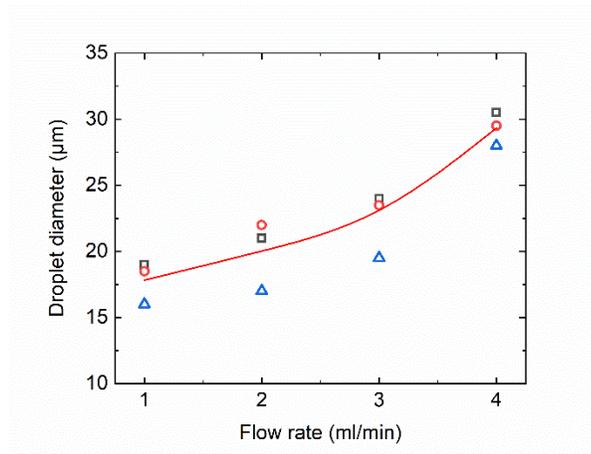

Fig. 10. Droplet diameter vs. volumetric ink flow rate from three independent runs; full (red) line is the average as a b-spline. The droplet diameter is increasing nonlinear as result of the increasing volumetric ink flow rate.

*Increasing atomizing power*

For completeness we studied the influence of the atomizing power from 1W to 4W at a spray height of 40mm, a flow rate of 1.5ml/min and a shroud pressure of 0.8psi as well and found only moderate influence on the droplet velocity and a linear increase in droplet diameter.

## 6. Conclusion

By a combination of PCS and Turbidimetry the determination of droplet diameter-velocity relations and their dependence on flow rate, shroud pressure and atomizing power of ultrasonically generated sprays was demonstrated. In our study, we show that this combination of PCS and Turbidimetry is a valuable solution to measure the droplet characteristics. To the best of our knowledge, measurements on droplets with PCS and Turbidimetry have not been explicitly demonstrated before.

The velocities measured with PCS are in agreement with the velocities measured with a High-Speed Camera. The measured droplet diameter confirms the theoretical predictions. Our modified setup was even able to characterize the droplet velocity and diameter more accurate and cost-efficient than using a High-Speed Camera where due to the high frame rates needed the spatial resolution is limited. This opens a way to map out the influences of the spray coat parameters on the quality of wetting related to the ink. This study extends the application and possibilities of PCS and Turbidimetry to droplets and the limits of particle size. Therefore, a combination of PCS and turbidimetry is a powerful tool to characterize ultrasonic generated droplets.

## 7. Disclosures

The authors declare no conflicts of interest.

## 8. Acknowledgement

The authors acknowledge the financial support by the DAAD through the project MPFL.

## 9. References

1.	D. Vak, S.-S. Kim, J. Jo, S.-H. Oh, S.-I. Na, J. Kim, and D.-Y. Kim, "Fabrication of organic bulk heterojunction solar cells by a spray deposition method for low-cost power generation," Applied Physics Letters **91**, 081102 (2007).
2.	C. N. Hoth, R. Steim, P. Schilinsky, S. A. Choulis, S. F. Tedde, O. Hayden, and C. J. Brabec, "Topographical and morphological aspects of spray coated organic photovoltaics," Organic Electronics **10**, 587-593 (2009).
3.	A. Abdellah, B. Fabel, P. Lugli, and G. Scarpa, "Spray deposition of organic semiconducting thin-films: Towards the fabrication of arbitrary shaped organic electronic devices," Organic Electronics **11**, 1031-1038 (2010).
4.	K. X. Steirer, M. O. Reese, B. L. Rupert, N. Kopidakis, D. C. Olson, R. T. Collins, and D. S. Ginley, "Ultrasonic spray deposition for production of organic solar cells," Solar Energy Materials and Solar Cells **93**, 447-453 (2009).
5.	K.-J. Kim, Y.-S. Kim, W.-S. Kang, B.-H. Kang, S.-H. Yeom, D.-E. Kim, J.-H. Kim, and S.-W. Kang, "Inspection of substrate-heated modified PEDOT: PSS morphology for all spray deposited organic photovoltaics," Solar Energy Materials and Solar Cells **94**, 1303-1306 (2010).
6.	S. F. Tedde, J. Kern, T. Sterzl, J. Furst, P. Lugli, and O. Hayden, "Fully spray coated organic photodiodes," Nano letters **9**, 980-983 (2009).
7.	K. Fujita, T. Ishikawa, and T. Tsutsui, "Novel method for polymer thin film preparation: spray deposition of highly diluted polymer solutions," Jpn. J. Appl. Phys. **41**, L70 (2002).
8.	T. Echigo, S. Naka, H. Okada, and H. Onnagawa, "Spray method for organic electroluminescent device fabrication," Jpn. J. Appl. Phys. **41**, 6219 (2002).
9.	G. S. Lonakar, M. S. Mahajan, S. S. Ghosh, and J. V. Sali, "Modeling thin film formation by Ultrasonic Spray method: A case of PEDOT: PSS thin films," Organic Electronics **13**, 2575-2581 (2012).
10.	H. Yi, J. Huang, X. Z. Gu, and Z. H. Ni, "Study on ultrasonic spray technology for the coating of vascular stent," Sci. China-Technol. Sci. **54**, 3358-3370 (2011).
11.	S.-I. Kwon, K.-H. Kyung, J.-Y. Park, I.-S. Lee, J.-H. Kim, S.-H. Kim, and S. Shiratori, "Uniform anti-reflective films fabricated by layer-by-layer ultrasonic spray method," Colloids and Surfaces A: Physicochemical and Engineering Aspects **580**, 123785 (2019).
12.	K. Gilissen, J. Stryckers, P. Verstappen, J. Drijkoningen, G. H. L. Heintges, L. Lutsen, J. Manca, W. Maes, and W. Deferme, "Ultrasonic spray coating as deposition technique for the light-emitting layer in polymer LEDs," Organic Electronics **20**, 31-35 (2015).
13.	E. J. Van Den Ham, S. Gielis, M. K. Van Bael, and A. Hardy, "Ultrasonic Spray Deposition of Metal Oxide Films on High Aspect Ratio Microstructures for Three-Dimensional All-Solid-State Li-ion Batteries," ACS Energy Letters **1**, 1184-1188 (2016).
14.	R. J. Lang, "ULTRASONIC ATOMIZATION OF LIQUIDS," J. Acoust. Soc. Am. **34**, 6-& (1962).
15.	J. Kim, K. Ha, and M. Kim, "Numerical analysis for the optimum condition of ultrasonic nebulizing," Jpn. J. Appl. Phys. **55**, 3 (2016).
16.	R. Rajan, and A. B. Pandit, "Correlations to predict droplet size in ultrasonic atomisation," Ultrasonics **39**, 235-255 (2001).

17. G. Cummins, and M. P. Y. Desmulliez, "Inkjet printing of conductive materials: a review," Circuit World **38**, 193-213 (2012).
18. C. Cornet, "Turbidimetry as a tool for the characterization of molecular-weight distributions," Polymer **9**, 7-14 (1968).
19. P. Walstra, "Estimating globule-size distribution of oil-in-water emulsions by spectroturbidimetry," Journal of Colloid and Interface Science **27**, 493-500 (1968).
20. G. E. Eliçabe, and L. H. García-Rubio, "Latex particle size distribution from turbidimetry using inversion techniques," Journal of colloid interface science **129**, 192-200 (1989).
21. T. D. Donnelly, J. Hogan, A. Mugler, M. Schubmehl, N. Schommer, A. J. Bernoff, S. Dasnurkar, and T. Ditmire, "Using ultrasonic atomization to produce an aerosol of micron-scale particles," Rev. Sci. Instrum. **76**, 10 (2005).
22. K. Gilissen, "Upscaling of Organic Light Emitting Devices: a focus on innivative materials, versatile printing techniques and state-od-the-art post treatments," in *Faculty of Enginieering Technology*(University Hasselt, 2016).
23. B. W. Mwakikunga, "Progress in Ultrasonic Spray Pyrolysis for Condensed Matter Sciences Developed From Ultrasonic Nebulization Theories Since Michael Faraday," Crit. Rev. Solid State Mat. Sci. **39**, 46-80 (2014).
24. W. Brown, *Light scattering: principles and development* (Oxford University Press, 1996).
25. B. Chu, *Laser light scattering: basic principles and practice* (Academic Press, 1974).
26. B. J. B. a. R. Pecora, *Dynamic Light scattering* (John Willey and sons, Inc, 1976).
27. W. Schärtl, "Light Scattering from Polymer Solutions and Nanoparticle Dispersions," in *Light Scattering from Polymer Solutions and Nanoparticle Dispersions*(springer, 2007).
28. G. Williams, and D. C. Watts, "Non-symmetrical dielectric relaxation behaviour arising from a simple empirical decay function," Transactions of the Faraday society **66**, 80-85 (1970).
29. R. Kohlrausch, "Theorie des elektrischen Rückstandes in der Leidener Flasche," Annalen der Physik **167**, 179-214 (1854).
30. C. Lindsey, and G. Patterson, "Detailed comparison of the Williams–Watts and Cole–Davidson functions," The Journal of chemical physics **73**, 3348-3357 (1980).
31. J. Rička, "Dynamic light scattering with single-mode and multimode receivers," Applied optics **32**, 2860-2875 (1993).
32. M. Plum, J. Rička, H. Butt, and W. Steffen, "Anisotropic hindered motion close to an interface studied by resonance-enhanced dynamic light scattering," New Journal of Physics **12**, 103022 (2010).
33. T. W. Chen, "Simple formula for light scattering by a large spherical dielectric," Applied optics **32**, 7568-7571 (1993).
34. T. W. Chen, "Diffraction by a spherical dielectric at large size parameter," Optics communications **107**, 189-192 (1994).
35. J. V. Champion, G. H. Meeten, and M. Senior, "Refraction by spherical colloid particles," Journal of Colloid and Interface Science **72**, 471-482 (1979).
36. A. G. Borovoi, "Light scattering by large particles: physical optics and the shadow-forming field," in *Light Scattering Reviews 8*(Springer, 2013), pp. 115-138.
37. K. Kraft, and A. Leipertz, "Sound velocity measurements by the use of dynamic light scattering: data reduction by the application of a Fourier transformation," Applied optics **32**, 3886-3893 (1993).
38. A. B. Leung, K. I. Suh, and R. R. Ansari, "Particle-size and velocity measurements in flowing conditions using dynamic light scattering," Applied optics **45**, 2186-2190 (2006).
39. T. Kourti, "Turbidimetry in Particle Size Analysis," (John Wiley & Sons, Ltd, 2006).
40. C. Dumouchel, P. Yongyingsakthavorn, and J. Cousin, "Light multiple scattering correction of laser-diffraction spray drop-size distribution measurements," International Journal of Multiphase Flow **35**, 277-287 (2009).
41. Ö. L. Gülder, "Multiple scattering effects in dense spray sizing by laser diffraction," Aerosol Science and Technology **12**, 570-577 (1990).